\documentclass[%
superscriptaddress,
nofootinbib,
amsmath,
amssymb,
prb,
twocolumn,longbibliography
]{revtex4-2}

\usepackage{dsfont}
\usepackage{bbm}
\usepackage{graphicx}
\usepackage{dcolumn}
\usepackage{bm}
\usepackage{xcolor}

\begin{document}

\title{Chirality Breaking of Majorana Edge Modes Induced by Chemical Potential Shifts}

\author{Xin Yue}
\email{Contact author: yuexin@csrc.ac.cn}
\affiliation{Beijing Computational Science Research Center, Beijing 100193, China}

\author{Guo-Jian Qiao}
\affiliation{Graduate School of China Academy of Engineering Physics, Beijing 100193, China}
\date{\today}

\begin{abstract}
Quantum anomalous Hall insulator-superconductor heterostructures are predicted to host chiral Majorana fermions as edge modes, which is essential for topological quantum computing applications. Although the edge states have been extensively studied at zero chemical potential $\mu = 0$, the practically relevant regime with a shifted chemical potential ($\mu \neq 0$) remains less explored. Here, we present an analytical treatment of the edge states for $\mu \neq 0$, deriving an approximate but highly accurate solution applicable to realistic experimental parameters. Surprisingly, we find that the energy dispersion of the edge band exhibits nonlinearity and transforms into a twisted, braid-like structure within specific parameter ranges. This unique braid-like band leads to non-chirality of the edge modes, allowing propagation in both directions.

\end{abstract}

\maketitle


\section{Introduction}
Majorana fermions, proposed in 1937 by Ettore Majorana, are exotic particles that are their own antiparticles \cite{majorana1937teoria}. In condensed matter physics, the quasi-excitations including Majorana zero modes and chiral Majorana fermion modes, have been extensively studied theoretically \cite{Read&Green2000,Kitaev_2001,Fu_liang_2008,Oreg_2010,Lutchyn_2010,Qi_2010,2010Proximity,Potter&Lee2011,Chung_2011,Alicea_2012,beenakker2013,Li2014,Wang_2015,Qiao_2022,Yue2023,LiuXin2024,Qiao2024, Legendre2024,Yue2025,qiao2025sizeoptimizationobserveingmajorana,zhang2025poor,2025Miao_braiding,APS2025,2025PRB_Osca, 2026PRB_WangJIng} and experimentally \cite{Mourik_2012,He2017chiral,Zhang2018,2020Absence,Uday2024,Huang_2024,2025PRB} due to their exotic properties and potential applications in fault-tolerant quantum computing \cite{Ivanov,Freedman2002,Kitaev_2003,Nayak2008,PNAS}. 
 
\begin{figure}
    \centering   \includegraphics[width=0.83\linewidth]{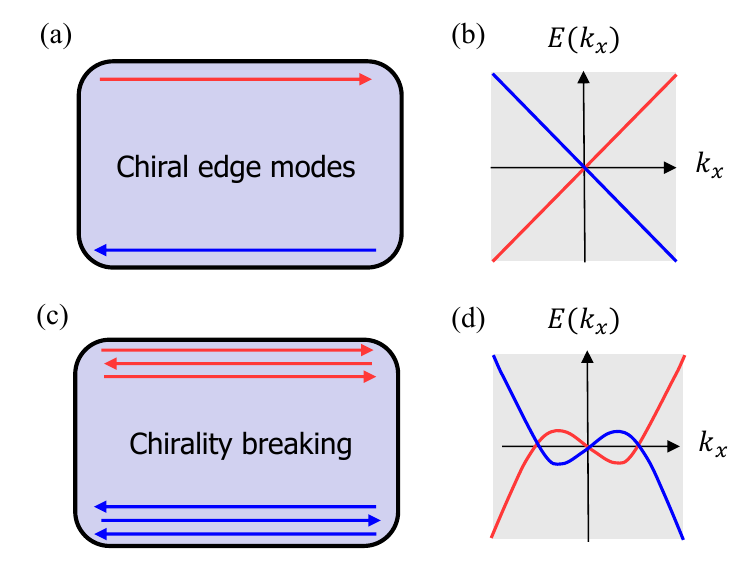}
    \caption{ (a) Illustration of chiral Majorana fermion propagating at the edge of 2D material, and (b) linear energy dispersion of the edge states. The 2D material could be 2D $p$-wave superconductor \cite{Read&Green2000}, or heterostructure form by QAH insulator and $s$-wave superconductor \cite{Qi_2010}. (c) Illustration for the chirality-breaking of Majorana edge modes under chemical potentail shift. (d) The dispersion relation of edge states shows a twisted, braid-like shape so that each edge band intersect Fermi level three times.}
    \label{chiral}
\end{figure}

The chiral Majorana fermion, as a quasiparticle, can be realized at the edge of 2D topological materials \cite{Read&Green2000}. Specifically, heterostructures \cite{Fu_liang_2008} composed of a quantum anomalous Hall (QAH) insulator and an $s$-wave superconductor are considered as a promising platform to host Majorana edge modes \cite{Qi_2010}. These modes propagate along the edges unidirectionally [see Fig. \ref{chiral}(a)], which is why they are referred to as chiral Majorana fermion modes or chiral Majorana edge modes \cite{Qi_2010,Chung_2011,Wang_2015,PNAS}.

These edge modes in heterostructure were studied in the special case where the effective chemical potential of the QAH insulator is zero, $\mu = 0$ \cite{Qi_2010,Yue2025}. In this scenario, the edge states exhibit a linear dispersion relation, as shown in Fig. \ref{chiral}(b). However, in practical experimental realizations, the chemical potential in the topological insulator layer can deviate by several meV  due to finite-size effect of the superconductor \cite{2017_Finite,Metallization2018,Metallization2022,2019_Awoga_Cayao,Qiao2024,Yue2025,qiao2025sizeoptimizationobserveingmajorana}. Therefore, it is necessary to study the edge states of this system when $\mu$ is not zero \cite{Legendre2024}.

In this paper, we investigate the edge states in the quantum anomalous Hall (QAH)-superconductor heterostructure for non-zero chemical potential ($\mu \neq 0$). While the formalism for analyzing edge states in topological systems is well-established \cite{Finite_edge2008,Anomalous_finite_edge2009,Shen_book,QSHE2008Review}, previous studies have typically involved solving quadratic equations to determine the decay length of edge states. The primary challenge for $\mu \neq 0$ is that the decay lengths are obtained by solving a quartic equation, which does not admit a simple analytical form. As a result, previous studies of similar systems have largely relied on numerical methods \cite{Legendre2024}. To overcome this, we present an analytical approach that yields highly accurate solutions across physically relevant parameter ranges, which are further validated by numerical calculations. Remarkably, the Majorana edge mode loses its chiral nature within specific regions of parameter space. In these regions, the energy dispersion of the edge states $E(k_x)$ becomes nonlinear and can exhibit a twisted, braid-like shape [see Fig. \ref{chiral}(d)]. This dispersion develops a kink such that the edge states intersect the Fermi level three times: twice with a positive group velocity and once with a negative group velocity \cite{Hasan&Kane_2010, twisting_edge2012} and consequently lose their chiral nature [see Fig. \ref{chiral}(c)]. The parameter region in which this braid-like edge state emerges is further determined using the analytical approach. Consequently, caution should be exercised when referring to “chiral Majorana fermions” at nonzero chemical potential.

\section{Edge States under Chemical Potential Shift}
The QAH-superconductor heterostructure can be described by a Bogoliubov–de Gennes Hamiltonian in momentum space \cite{Qi_2010,Chung_2011}:
\begin{equation}
\mathcal{H}(\bm{k}) = 
\begin{pmatrix}
\bm{h}_{Q}(\bm{k}) - \mu & \mathrm{i}\Delta\bm{\sigma}_y \\
-\mathrm{i}\Delta\bm{\sigma}_y & -\bm{h}_{Q}^*(-\bm{k}) + \mu
\end{pmatrix},
\end{equation}
where the Hamiltonian of the QAH insulator is \cite{RuiYu_2010}
\begin{equation}
\bm{h}_{Q}(\bm{k}) = A(k_x\bm{\sigma}_x + k_y\bm{\sigma}_y) + \left(m - B k^2\right)\bm{\sigma}_z,
\end{equation}
and $\Delta$ denotes the superconducting pairing in the QAH insulator induced by the proximity effect. Here, $\bm{k} = (k_x, k_y)$ is the electron momentum, $\mu$ is the chemical potential, $\bm{\sigma}_i$ ($i = x, y, z$) is the Pauli matrices, and $A$,$B$ and $m$ are material-specific parameters.

We consider a semi-infinite plane ($y \ge 0$) with open boundary conditions along the $y$ direction and translational invariance along $x$. Under this geometry, $k_x$ remains a good quantum number, denoted as $k_x$, while $k_y$ is replaced by the differential operator $-\mathrm{i}\partial_y$. Then the eigenvalue equation becomes
\begin{equation}
\mathcal{H}\bigl(k_x, -\mathrm{i}\partial_y\bigr)\, \bm{\Psi}_{k_x}(y) = E\, \bm{\Psi}_{k_x}(y).
\end{equation}
To search for zero-energy modes localized at the edge $y=0$, we first focus on the $k_x=0$ and $E=0$ case. A localized ansatz can be written as
\begin{equation}
\bm{\Psi}_{k_x}(y) = \mathrm{e}^{-\xi y} \begin{pmatrix} \bm{u}_{k_x} \\ \bm{v}_{k_x} \end{pmatrix},
\end{equation}
where $\operatorname{Re}(\xi)>0$ ensures decay as $y\rightarrow\infty$. In this section we only focus on the  $\mathcal{N}= 1$  topological superconducting case.

\subsection{Decay lengths and Wave Functions}

The decay factors $\xi$ are determined by the characteristic equation $\det\!\bigl[\mathcal{H}(0,\mathrm{i}\xi)\bigr] = 0$, whose left-hand side factorizes into a product of two quartic polynomials $F_1(\xi)F_2(\xi)$ with
\[
F_{1,2}(\xi) = B^2\xi^4 - \bigl(2Bm+A^2\bigr)\xi^2 \mp 2A\Delta\xi + m^2 - \mu^2 - \Delta^2.
\]
Each quartic polynomial yields four algebraic roots for $\xi$, so their product generates a full set of eight decay-length solutions total. In our physical parameter regime, the quadratic kinetic coefficient $B$ takes a numerically small value satisfying the uniform small-$B$ bound $B \ll {A^2}/{X}$ with
\[
\qquad X=\max\bigl(\Delta,|m|,\sqrt{|m^2 - \mu^2 - \Delta^2|},\Delta \pm \sqrt{m^2 - \mu^2}\bigr),
\]
which permits a controlled perturbative splitting of the full root set into two well-separated families with vastly different magnitudes. When restricting the analysis to the case $k_x=0$ and $E=0$, the  system can be equivalently mapped onto the one-dimensional nanowire-superconductor model \cite{Oreg_2010,Lutchyn_2010}, where a similar quartic equation emerges. We summarize the leading-order asymptotic approximations valid under this constraint, with full algebraic derivation and validity justification collected in Appendix A.

\textit{Four small roots} --
When we treat $B$ as a subleading small parameter and neglect all $B$-dependent terms in the quartic polynomials $F_{1,2}(\xi)$, the resulting quadratic algebraic equations produce four finite low-magnitude solutions
\begin{align}
\xi_{1,2} &\approx \frac{\Delta \pm \sqrt{m^2 - \mu^2}}{A}, 
\xi_{3,4} \approx -\frac{\Delta \pm \sqrt{m^2 - \mu^2}}{A}.
\end{align}
Finite non-zero $B$ only introduces minor quantitative corrections to these roots.

\textit{Four large roots} --
The remaining four solutions arise from balancing the dominant $B^2\xi^4$ and $A^2\xi^2$ terms in the full quartic polynomials, retaining only the highest-order powers of $\xi$ for leading-order analysis. This balance yields roots scaling as $\xi \propto 1/B$,
\begin{equation}
\xi_{5,6} \approx \pm\frac{A}{B},\qquad \xi_{7,8} \approx \pm\frac{A}{B}.
\end{equation}
Since $B$ is small and obeys $B\ll A^2/X$, these roots are orders of magnitude larger than the four roots obtained above.

The corresponding un-normalized wave functions are:
\begin{align*}
\bm{\Psi}_1 &= \mathrm{e}^{-\xi_1 y}(\sigma,-1,-\sigma,1)^{\mathrm{T}},\;
 &\bm{\Psi}_2 = \mathrm{e}^{-\xi_2 y}(\sigma,1,-\sigma,-1)^{\mathrm{T}},\\
\bm{\Psi}_3 &= \mathrm{e}^{-\xi_3 y}(\sigma,1,\sigma,1)^{\mathrm{T}},\;
 &\bm{\Psi}_4 = \mathrm{e}^{-\xi_4 y}(\sigma,-1,\sigma,-1)^{\mathrm{T}}.\\
\bm{\Psi}_5 &= \mathrm{e}^{-\xi_5 y}(1,1,1,1)^{\mathrm{T}},\;
 &\bm{\Psi}_6 = \mathrm{e}^{-\xi_6 y}(1,-1,1,-1)^{\mathrm{T}},\\
\bm{\Psi}_7 &= \mathrm{e}^{-\xi_7 y}(1,1,-1,-1)^{\mathrm{T}},\;
 &\bm{\Psi}_8 = \mathrm{e}^{-\xi_8 y}(1,-1,-1,1)^{\mathrm{T}},
\end{align*}
where $\sigma \equiv (m+\mu)/\sqrt{m^2-\mu^2}$.

\subsection{Boundary Conditions and Eigenstates}

The edge states must satisfy the boundary conditions $\bm{\Psi}(y=0) = 0$ and $\bm{\Psi}(y \to \infty) = 0$. Therefore, only solutions with $\operatorname{Re}(\xi) > 0$ are physically admissible when considering edge modes localized near $\bm{\Psi}(y=0) = 0$. Furthermore, to satisfy the boundary condition at $y=0$, the edge states must be constructed as a linear combination of these admissible solutions. 

In the parameter regime $|m| < \sqrt{\Delta^2+\mu^2} $, with $A, B > 0$, it follows that $\operatorname{Re}(\xi_{1}),\operatorname{Re}(\xi_{2}) > 0$ and $\xi_5=\xi_7 = A / B > 0$, while the remaining roots possess a negative real part and correspond to non-physical growing modes. Consequently, a trial wavefunction can be written as:
\begin{equation}
\bm{\Psi}(y) = c_1 \bm{\Psi}_1(y) + c_2 \bm{\Psi}_2(y) + c_5 \bm{\Psi}_5(y)+c_7 \bm{\Psi}_7(y),
\end{equation}
where $\bm{\Psi}_{1,2,5,7}(y)$ are eigenstate solutions associated with the positive roots. The boundary condition $\bm{\Psi}(0) = 0$ leads to the following set of equations:
\begin{align*}
(c_1 + c_2)\sigma + c_5 + c_7 &= 0, & -c_1 + c_2 + c_5 + c_7 &= 0,  \\
-(c_1 + c_2)\sigma + c_5 - c_7 &= 0, & c_1 - c_2 + c_5 - c_7 &= 0. 
\end{align*}
By setting $c_7 = -1$, one obtains:
\begin{equation}
c_1 = \frac{1-\sigma}{2\sigma}, \qquad c_2 = \frac{1+\sigma}{2\sigma}, \qquad c_5=0.
\end{equation}
It is straightforward to verify from Eqs.(7) and (8) that the electron and hole components of the wave function satisfy $\Psi_{e \uparrow} = \Psi_{h \uparrow}$. This property originates from the particle-hole symmetry of the system and indicates that the edge mode is a Majorana-type mode.

\subsection{Edge States in Different Chemical Potential}

\begin{figure}
    \centering
    \includegraphics[width=1\linewidth]{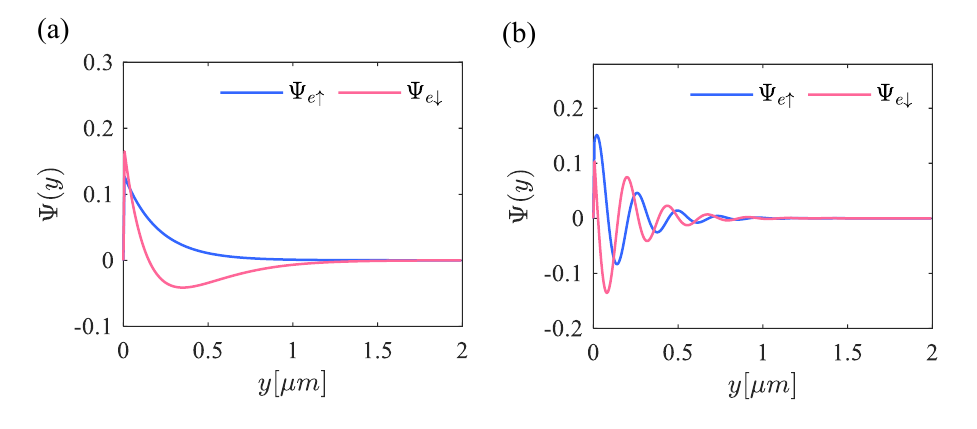}
    \caption{Edge state wave function under a chemical potential shift. Only the electron component is displayed, as the hole component is identical to the electron component. (a) $|\mu|=|m|$, where the spin-up component of the edge state $\bm{\Psi}_{e\uparrow} $ decays exponentially, while the spin-down component $\bm{\Psi}_{e\downarrow}$  exhibits an exponential decay modified by a linear factor $(1+\frac{2m}{A} y)$. (b) $|\mu|>|m|$, where the edge states oscillate and decay simultaneously.}
    \label{Fig2}
\end{figure}

In the above analysis, the general form of the edge state wavefunction was derived under the conditions $|m| < \sqrt{\Delta^2+\mu^2}$ and $k_x = 0$. In this section, we present explicit expressions for the edge states in different chemical potential regimes.

(i) For $|\mu| < |m|$, the electron components of the edge-state wave function are given by: 
\begin{align*}
\Psi_{e \uparrow} &= 
\frac{1+p}{2}
\mathrm{e}^{-\frac{\Delta - S}{A}y} +\frac{1-p}{2}\mathrm{e}^{-\frac{\Delta + S}{A}y} - \mathrm{e}^{-\frac{ A}{B}y}, \\
\Psi_{e \downarrow} &= \frac{1+p}{2p}
\mathrm{e}^{-\frac{\Delta - S}{A}y} +\frac{1-p}{2p}\mathrm{e}^{-\frac{\Delta + S}{A}y} - \mathrm{e}^{-\frac{ A}{B}y},
 \end{align*}
with $p\equiv (m+\mu)/S $ and $\  S \equiv \sqrt{m^2 - \mu^2}$. 

(ii) For $|\mu| > |m|$, the edge state wave function exhibits oscillatory decay, as shown in Fig. \ref{Fig2}(b). Specifically, one finds:
\begin{align*}
\Psi_{e \uparrow} &= \cos\left(\frac{Q y}{A}\right) \mathrm{e}^{-\frac{\Delta }{A}y}  + \frac{m + \mu  }{Q} \sin\left(\frac{Q y}{A}\right)\mathrm{e}^{-\frac{\Delta }{A}y}- \mathrm{e}^{-\frac{ A}{B}y}, \\
\Psi_{e \downarrow} &= \cos\left(\frac{Q y}{A}\right) \mathrm{e}^{-\frac{\Delta }{A}y} - \frac{Q }{m + \mu} \sin\left(\frac{Q y}{A}\right) \mathrm{e}^{-\frac{\Delta }{A}y} - \mathrm{e}^{-\frac{ A}{B}y},
\end{align*}
where $Q \equiv \sqrt{\mu^2 - m^2}$. These solutions describe oscillatory Majorana edge states in the regime where the chemical potential exceeds the effective mass gap.

(iii) When $\mu = -m$, a limiting procedure can be employed to derive the explicit wavefunction forms. The resulting components are:
\begin{align}
\Psi_{e \uparrow} &= \mathrm{e}^{-\frac{\Delta}{A} y} - \mathrm{e}^{-\frac{A}{B} y}, \\
\Psi_{e \downarrow} &= \left( 1 + \frac{2m}{A} y \right)\mathrm{e}^{-\frac{\Delta}{A} y}  - \mathrm{e}^{-\frac{A}{B} y}.
\end{align}
In this case, both the spin-up and spin-down components of the edge state decay exponentially [see Fig. 2(a)].

(iv) When $\mu = 0$ and \(m < 0\), the wavefunction $\bm{\Psi}$ simplifies to a linear combination of two basis states, consistent with the edge-state solutions commonly presented in the literature for systems such as the quantum spin Hall insulator \cite{Finite_edge2008, Shan_2010, QSHE2008Review}, topological insulator $\rm Bi_{2}Se_{3}$ \cite{Anomalous_finite_edge2009}
 and the lattice model of a 1D \(p\)-wave superconductor \cite{Kitaev_2001}. Specifically:
\begin{equation}
\bm{\Psi}_{e\uparrow} =     \mathrm{e}^{-\frac{y (\Delta - m)}{A}} - \mathrm{e}^{-\frac{y A}{B}}.
\end{equation}

These expressions demonstrate the distinctive decay
behavior of the edge states in different chemical potential shift regime. It should be noted that all analytical formulas herein are derived under the condition $|m| < \sqrt{\Delta^2+\mu^2}$, corresponding to the $\mathcal{N}=1$ phase.

\section{Energy band structure under chemical potential shift}
Due to the presence of proximity-induced pairing in the QAH system, the bands do not simply shift uniformly when the chemical potential $\mu$ is varied, particularly for the edge states. To explore this behavior quantitatively, we map the continuous model onto a discrete lattice. The energy spectra are obtained by numerically diagonalizing the Hamiltonian \cite{Qi_2010,QSHE2008Review,Shen_book}, with open boundary conditions along the $y$-direction and periodic boundary conditions along the $x$-direction.

\begin{figure}
    \centering
    \includegraphics[width=1\linewidth]{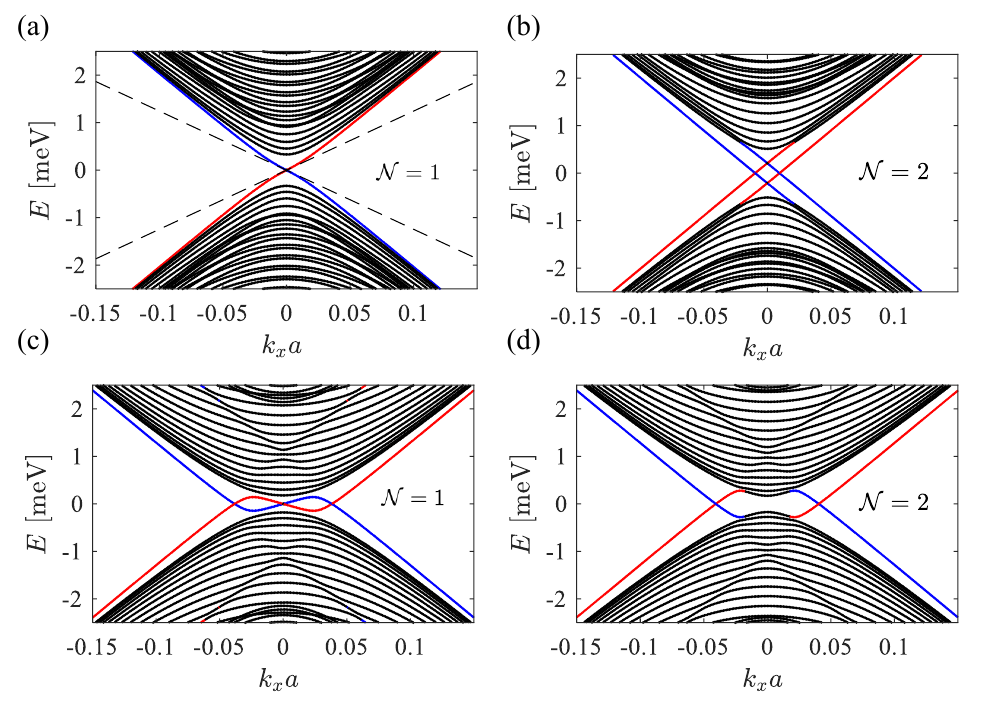}
   \caption{Energy spectra of the heterostructure system with various parameters. The edge bands are highlighted in red and blue. (a) $m=-0.6 \, \mathrm{meV}$, $\mu=1 \, \mathrm{meV}$. (b) $m=2 \, \mathrm{meV}$, $\mu=0.5 \, \mathrm{meV}$. (c) $m=-2 \, \mathrm{meV}$, $\mu=2 \, \mathrm{meV}$, where the edge bands form a braid-like structure. (d) $m=-2.5 \, \mathrm{meV}$, $\mu=2 \, \mathrm{meV}$. Other parameters are identical for (a--d): $\Delta_s = 1 \, \mathrm{meV}$,$A = 0.3 \, \mathrm{meV}\cdot\mu\mathrm{m}$, and $B = 1.5 \times 10^{-4} \, \mathrm{meV}\cdot\mu\mathrm{m}^2$, with a lattice constant $a = 5 \, \mathrm{nm}$ and a length $L_y = N_y a = 2 \, \mu\mathrm{m}$ along the $y$ direction.}

    \label{fig:fig3}
\end{figure}

Under these boundary conditions, the wave function of the bulk states generally takes the form \(\psi \sim \exp(\mathrm{i} k_x x) \sin(k_y y)\), while the wave function of the edge states generally takes the form \(\psi \sim \exp(\mathrm{i} k_x x) \exp(-\xi y)\). Therefore, the edge and bulk states can be distinguished by analyzing the spatial distribution of the eigenstates. Specifically, if the majority of the wave function (set to 80\%) is localized on the $y=0$ boundary, it is identified as a edge state and marked with red dots in Fig. 3. If the majority of the wave function is localized on the $y=L_y$ boundary, it is identified as a edge state and marked with blue dots. If neither condition is satisfied, the state is considered a bulk state and marked with black dots (see Fig.~\ref{fig:fig3}).

\begin{figure*}[htb]
    \centering
    \includegraphics[width=0.85\linewidth]{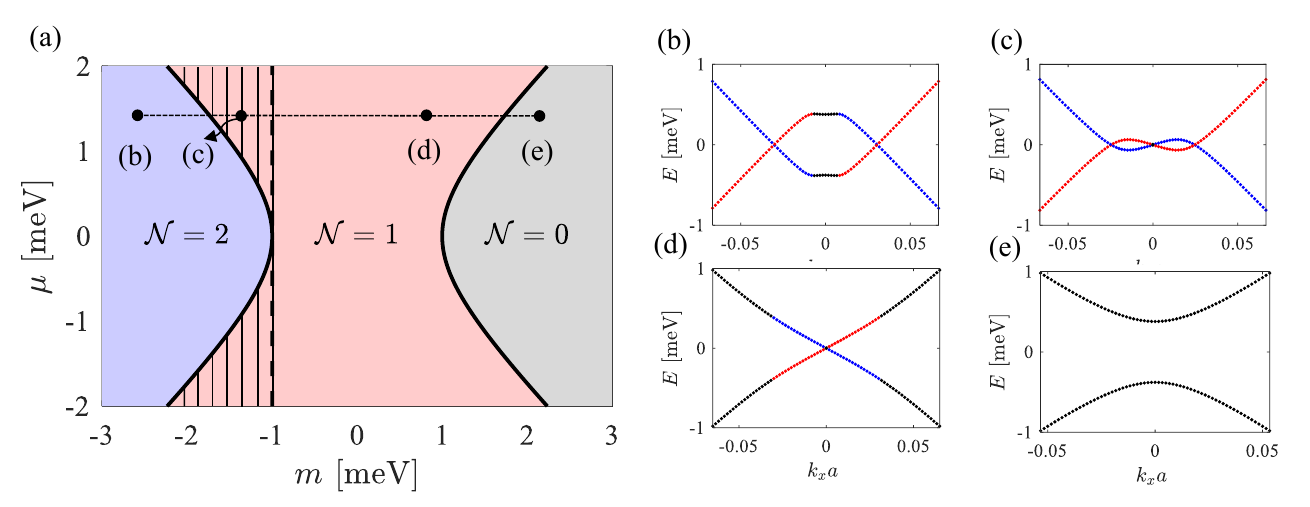}
    \caption{This figure illustrates the topological phase diagram as a function of \(m\) and \(\mu\), with the shaded regions representing areas where braid-like edge bands emerge. Panels (b)-(e) correspond to energy spectra at specific parameter sets labeled in (a), only the lowest band is shown for clarity. Other parameter are setting the same as that in Fig. 3.}
    \label{fig:fig5}
\end{figure*}

For the $\mathcal{N}= 2$ topological phase (defined by $m < -\sqrt{\Delta^2 + \mu^2}$) \cite{Qi_2010,Chung_2011}, we observe two edge bands on each side. These edge bands shift collectively as the chemical potential varies. In contrast, for the $\mathcal{N}= 1$ topological phase (defined by $|m| < \sqrt{\Delta^2+\mu^2}$) \cite{Qi_2010,Chung_2011}, only one edge band exists on each side, and its position is not affected by changes in the chemical potential. Interestingly, we find that in certain parameter regimes, the edge bands form a braid-like structure, as shown in Fig.~\ref{fig:fig3}(c). A natural question is what parameter range gives rise to this braid-like dispersion of the edge state.

\section{Conditions for the emergence of braid-like bands}

As shown above, the dispersion of the edge states can exhibit a braid-like structure under specific parameter settings. To determine the origin of these exotic structure, we examine the edge-state dispersion under perturbation by small $k_x$. For $k_x \neq 0$, the Hamiltonian can be decomposed as 
\begin{equation}
    \mathcal{H}(k_x, -\mathrm{i}\partial_y) = \mathcal{H}_0(-\mathrm{i}\partial_y) + \mathcal{H}_1(k_x),
\end{equation}
where the perturbative term $\mathcal{H}_1(k_x)$ is expressed as
\begin{equation}\label{eq:H1kx}
    \mathcal{H}_1(k_x) =
    \begin{pmatrix}
        B k_x^2 & A k_x & 0 & 0 \\
        A k_x & -B k_x^2 & 0 & 0 \\
        0 & 0 & -B k_x^2 & A k_x \\
        0 & 0 & A k_x & B k_x^2
    \end{pmatrix}.
\end{equation}
To first order in $k_x$, the group velocity $v$ of edge modes can be derived from standard perturbation theory. Using the wave function $\bm{\Psi}(y) = c_1 \bm{v}_1 \mathrm{e}^{-\xi_1 y} + c_2 \bm{v}_2 \mathrm{e}^{-\xi_2 y} + c_7 \bm{v}_7 \mathrm{e}^{-\xi_7 y}$ for edge states, the dispersion energy $E(k_x)$ is given by
\begin{align}
    E(k_x) &= \int_0^\infty \mathrm{d}y \, \bm{\Psi}^\dagger(y) \mathcal{H}_1(k_x) \bm{\Psi}(y) \nonumber \\
    &= \sum_{i,j=1,2,7} c_i^* c_j 
    \left(\bm{v}_i^\dagger \mathcal{H}_1(k_x) \bm{v}_j\right)
    I_{ij},
\end{align}
where $\bm{v}_{i}=\bm{\Psi}_i\mathrm{e}^{\xi_iy}$ and $I_{ij} = \int_0^\infty \mathrm{e}^{-\xi_i^*y} \mathrm{e}^{-\xi_jy} \mathrm{d}y =1/(\xi_i^* + \xi_j)$ represents overlap integrals stemming from the edge-state wave functions. Since $B$ is typically small for realistic materials, terms involving $\bm{v}_7$ (corresponding to $\xi_7 = A/B \gg \xi_1, \xi_2$) can be neglected. This greatly simplifies the result to the following dispersion relation [see Appendix B for calculation details]:
\begin{equation}
    E(k_x) = \frac{\Delta^2 + m \Delta}{\Delta^2 + m \Delta + \mu^2} A k_x.
    \label{dispersion}
\end{equation}
This analytical expression is consistent with numerical results obtained from diagonalizing the lattice model. For $\mu = 0$ (corresponding to $\sigma = 1$), the results reproducing the linear dispersion $Ak_x$ reported widely in studies of topological edge states \cite{Qi_2010,Shen_book,2017EPL}.

It is worth noting that the dispersion relation Eq. (\ref{dispersion}) is only applicable in the topological region $\mathcal{N}=1$, since our calculation starts from the eigenvalue $E(k_x=0)=0$. This eigenvalue ceases to exist when $\mathcal{N}=2$, and therefore the above calculation does not apply. In the $\mathcal{N}=2$ case, the two edge bands at one side shift by the chemical potential as $E = A k_x \pm \mu$ in the perturbative sense. When the system undergoes a phase transition from $\mathcal{N}=2$ to $\mathcal{N}=1$, the two separated shifted edge bands fuse into a single band, forming an \textit{N}-shaped curve with linear asymptotes. The \textit{N}-shaped curves (red) and inverted N-shaped curves (blue) intertwine with each other, resembling a braid [see Fig. 4(c)]. 

The critical point where the \textit{N}-shaped curve appears corresponds to where the slope of the dispersion at $k_x = 0$ vanishes. It follows from Eq.~(\ref{dispersion}) that the critical condition is $\Delta + m = 0$. When $\Delta + m < 0$, the numerator of the dispersion relation becomes negative, while the denominator is always positive in the $\mathcal{N}=1$ region. Therefore, the region defined by $\Delta + m < 0$ and $m > -\sqrt{\Delta^2 + \mu^2}$ corresponds to the parameter regime where the braid-like edge bands emerge, as illustrated by the shaded region in Fig. 4(a).

\section{Non-Chirality in the braid-like band}
\subsection{What is chirality?}
Chirality in edge transport refers to the presence of modes that propagate only in one direction without counter-propagating partners. A paradigmatic example can be found in the edge mode of a quantum Hall state \cite{Halperin1982,Hasan&Kane_2010}, a quantum anomalous Hall state \cite{Haldane1988,RuiYu_2010} or a two-dimensional \(p\)-wave superconductor \cite{Read&Green2000}, characterized by the linear dispersion \(E(k_x) = A k_x\). The group velocity \(v_g = \partial E / \partial k_x = A\) is constant, indicating that all excitations propagate uniformly in the same direction.

To explicitly describe the motion of the Majorana edge modes, we consider an initial wave packet composed of edge eigenstates localized at the boundary, with momentum components drawn from the dispersion relation:
\begin{equation}
\bm{\Psi}(x,y) = \int \mathrm{d}k_x\, g(k_x) \mathrm{e}^{\mathrm{i}k_{x}x} \bm{\Psi}_{k_x}(y),
\end{equation}
where \(g(k_x)\) is a Gaussian distribution centered at a chosen \(k_0\), and \(\bm{\Psi}_{k_x}(y)\) are the edge states. During time evolution, the group velocity \(v_g(k_x) = \mathrm{d}E/\mathrm{d}k_x\) governs the propagation speed of the wavepacket center \cite{PNAS,dong2025,2025manipulation_MWP}. For a band with linear dispersion  [see Fig. \ref{chiral}(b)], the sign of \(v_g\) is fixed as $k_x$ varies, ensuring the wavepacket propagates in a single direction.

\subsection{Why are the braid-like edge bands non-chiral?}
The braid-like edge bands discussed previously exhibit a distinctly different behavior due to their nonlinear dispersion structure. In such bands, the group velocity \(v_g(k_x) = \mathrm{d}E/\mathrm{d}k_x\) becomes a non-monotonic function, changing sign at certain points within the Brillouin zone [see Fig. \ref{chiral}(d)].

This non-monotonicity has profound consequences for wavepacket propagation. As the wavepacket evolves, portions corresponding to positive slopes propagate forward (right-moving), while portions corresponding to negative slopes propagate backward (left-moving). This division results in the splitting of the wavepacket, rather than a unidirectional propagation. Furthermore, at the momentum points where \(\mathrm{d}E/\mathrm{d}k_x = 0\), the group velocity vanishes, causing the center of the wavepacket to hold during its propagation.

\subsection{What can we do with these non-chiral properties?}

The most striking consequence of the braid-like edge bands is the non-monotonicity of the group velocity \(v_g(k_x) = \partial E/\partial k_x\), which undergoes a sign reversal at specific momentum points. Notably, this group velocity sign reversal enables precise control over the propagation dynamics of Majorana edge-mode wavepackets. By dynamically tuning the system parameters (e.g., chemical potential \(\mu\), effective mass \(m\), or induced superconducting pairing \(\Delta\)), we can manipulate the dispersion profile of the braid-like bands, thereby regulating the group velocity of the wavepacket. Specifically, we can engineer three key propagation behaviors: (1) positive group velocity, leading to forward (right-moving) propagation of the wavepacket; (2) zero group velocity, achieved at the momentum points where \(v_g(k_x) = 0\), resulting in the arrest of wavepacket propagation; and (3) negative group velocity, inducing backward (left-moving) propagation. 

Nevertheless, such flexible velocity manipulation comes with an inherent physical trade-off. The braid-like edge dispersion intrinsically exhibits momentum dispersion, namely group velocity varies continuously with \(k_x\). A physical Majorana wavepacket consists of a distribution of \(k_x\) components, each propagating at a distinct group velocity. This mismatch inevitably leads to gradual wavepacket broadening during dynamical evolution. To achieve effective and coherent wavepacket manipulation, the system parameters must be carefully tuned such that the characteristic time scale of wavepacket diffusion is substantially longer than the propagation time scale of the edge mode. 

\section{Comparison between chiral symmetry breaking and chirality breaking}
In this section, we distinguish the chiral symmetry breaking discussed in this work from the related concepts reported in previous literature \cite{2012PRL_Topo,2012PRBchiral_symmetry,2018PRB_quasione}. In general, the term \textit{chirality breaking} describes the loss of unidirectional propagation for edge states. In contrast, \textit{chiral symmetry breaking} refers to the breakdown of the inherent symmetry of the system’s Hamiltonian. These two physical phenomena are conceptually distinct and independent of each other.

The Hamiltonian of the studied system possesses three fundamental symmetries. The chiral symmetry is defined as
\begin{equation}
\mathcal{C} \mathcal{H}(k_x) \mathcal{C}^{-1} = -\mathcal{H}(k_x),
\end{equation}
where the chiral operator is constructed as $\mathcal{C} = \mathcal{P}\mathcal{T}$ \cite{2018PRB_quasione}. Here, $\mathcal{P} = \bm{\tau}_x \mathcal{K}$ denotes the particle-hole symmetry operator, $\mathcal{T}$ is the time-reversal symmetry operator, $\bm{\tau}_x$ is the Pauli matrix acting on the electron-hole degree of freedom, and $\mathcal{K}$ represents the complex conjugation operator. For a pristine system without disorder, chiral symmetry is always preserved, regardless of whether the edge states exhibit chiral propagation characteristics.

We further investigate the robustness of the braid-like edge bands against disorder. When random disorder is introduced, the chiral symmetry of the system is broken, which modifies the morphology of edge bands. Nevertheless, the particle-hole symmetry remains intact, imposing strict constraints on the band structure. The spectral constraints originating from the three symmetries are summarized as follows:
(i) From particle-hole symmetry: If $E(k_x)$ is an eigenvalue of $\mathcal{H}(k_x)$ with corresponding eigenstate $\psi_{k_x}$, then $-E(-k_x)$ is also an eigenvalue associated with the eigenstate $\mathcal{P}\psi_{-k_x}$.
(ii) From time-reversal symmetry: If $E(k_x)$ is an eigenvalue of $\mathcal{H}(k_x)$ with eigenstate $\psi_{k_x}$, then $E(-k_x)$ is an eigenvalue with eigenstate $\mathcal{T}\psi_{-k_x}$.
(iii) From chiral symmetry: If $E(k_x)$ is an eigenvalue of $\mathcal{H}(k_x)$ with eigenstate $\psi_{k_x}$, then $-E(k_x)$ is an eigenvalue with eigenstate $\mathcal{C}\psi_{k_x}$.

\begin{figure}
    \centering
    \includegraphics[width=1\linewidth]{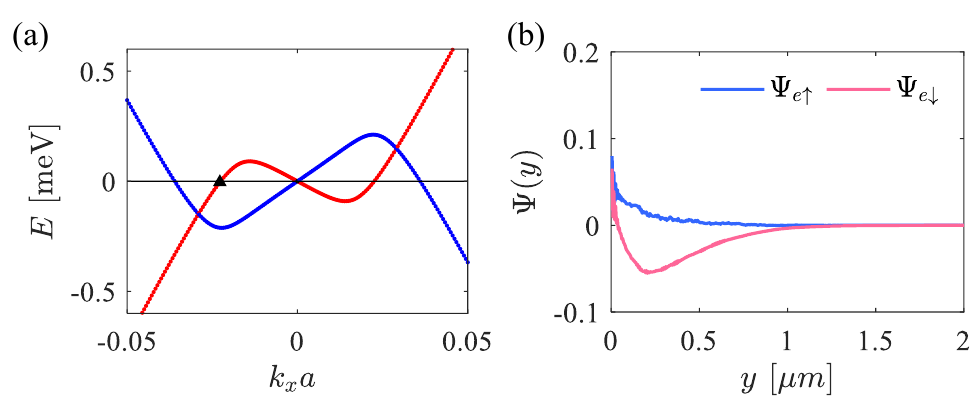}
    \caption{
    (a) Edge band structure under on-site disorder with disorder strength $W=2\,\mathrm{meV}$. All structural parameters are identical to those adopted in Fig.~3(c). Particle-hole symmetry remains intact, evidenced by the spectral pairing relation $E(k_x) \leftrightarrow -E(-k_x)$. In contrast, chiral symmetry is broken, which lifts the constraint $E(k_x) \leftrightarrow -E(k_x)$ for energy eigenvalues. (b) Edge-state wave function evaluated at the black triangle labeled in panel (a), corresponding to $k_xa=-0.023$ and $E=0$.
    }
    \label{fig:disorder}
\end{figure}

In our numerical simulations, we introduce on-site disorder in the form of random potential shifts at each lattice site along the $y$ direction. The disorder potential at each site is expressed as
\begin{equation}
V_j = W \cdot \big(\xi_j - 0.5\big),
\end{equation}
where $j$ labels lattice sites, $W$ is the disorder strength, and $\xi_j \in [0,1]$ denotes independent uniform random numbers. The resultant edge band structure under disorder strength $W=2\,\mathrm{meV}$ is presented in Fig.~\ref{fig:disorder}, with all other system parameters identical to those used in Fig.~3(c).

The numerical results clearly verify our symmetry analysis. The particle-hole symmetry survives disorder, such that each energy level $E(k_x)$ always has a counterpart $-E(-k_x)$ in the spectrum. By contrast, chiral symmetry is broken by disorder, so the energy levels no longer satisfy the pairwise relation $E(k_x) \leftrightarrow -E(k_x)$. Although the disorder distorts the edge-state structure and lifts the band crossing at $E=0$, the overall braid-like profile of the edge bands is well retained. We therefore conclude that the braid-like feature of edge bands is robust against moderate disorder.

It is worth noting that the disorder considered in this study is applied only along the $y$-direction. The robustness of braid-like modes under more realistic disorder conditions, such as uncorrelated and correlated puddle-type disorder~\cite{Kristof_2025} under fully open boundary conditions in both the $x$ and $y$ directions, deserves further investigation.

\section{Conclusion and Outlook}

In this study, we investigate the edge states of quantum anomalous Hall insulator-superconductor heterostructures under a non-zero chemical potential (\(\mu \neq 0\)), addressing a critical gap between theoretical predictions at \(\mu = 0\) and practical experimental conditions where chemical potential shifts are unavoidable due to finite-size effects and interface interactions. By solving the quartic characteristic equation derived from the Bogoliubov–de Gennes Hamiltonian, we obtain approximate but highly accurate analytical solutions for the edge-state wavefunctions and their decay lengths, which are validated by numerical simulations of the lattice model. Our key finding is that the chiral Majorana edge modes, which exhibit linear dispersion and unidirectional propagation at \(\mu = 0\), undergo a profound chirality-breaking transition under non-zero chemical potential in specific parameter regimes.

We demonstrate that when the system parameters satisfy \(\Delta + m < 0\) and \(m > -\sqrt{\Delta^2 + \mu^2}\) (within the \(\mathcal{N}=1\) topological phase), the edge-state energy dispersion transforms into a distinctive braid-like structure. Our analytical treatment of the edge-state dispersion relation, derived via first-order perturbation theory, provides a quantitative framework for understanding the dependence of edge-mode propagation on chemical potential and other material parameters. These results establish that chemical potential shifts are a powerful knob for tuning the band structure of Majorana edge modes, refining the conventional understanding of their inherent chirality and
 enabling flexible dynamical manipulation of these exotic quasiparticle modes.

\begin{acknowledgments}
The authors would like to thank C. P. Sun for originally stimulating their interest in this problem and for many illuminating discussions. This study was supported by the Science Challenge Project (Grant No.TZ2025017), the National Natural Science Foundation of China (NSFC) (Grant No. 12088101, 12547124), and the China Postdoctoral Science Foundation (Grant No. 2025M784438).
\end{acknowledgments}

\section*{Note added}
Upon completion of our manuscript, we became aware of a recent paper \cite{reich2026_Karlsruhe}, which observed an analogous braid-like dispersion of one-dimensional bound states in Josephson junctions formed on the surface of a three-dimensional topological insulator under large chemical potentials. This cross-platform correspondence strongly suggests that the chirality-breaking mechanism we uncover is a general phenomenon in topological systems, rather than a system-specific effect, thus highlighting the broader relevance of our findings to the field of topological matter.

\section*{data availability}
The data that support the findings of this article are not publicly available. The data are available from the authors upon reasonable request.

 \appendix 

 \onecolumngrid

\section{Detailed Derivation of Decay lengths}
The BdG matrix for the $k_x=0$ case with $k_y\rightarrow \mathrm{i}\xi$ reads:
\begin{align}
\mathcal{H}(0,\mathrm{i}\xi) = \begin{pmatrix}
-B\xi^2+m-\mu & A\xi & 0 & \Delta \\
-A\xi & B\xi^2-m-\mu & -\Delta & 0 \\
0 & -\Delta & B\xi^2-m+\mu & -A\xi \\
\Delta & 0 & A\xi & -B\xi^2+m+\mu
\end{pmatrix}.
\end{align}
The determinant factorization gives: $\det\!\bigl[\mathcal{H}(0,\mathrm{i}\xi)\bigr]= F_1(\xi)F_2(\xi)$, where
\begin{align}
F_1(\xi) &= B^2\xi^4 - 2Bm\xi^2 - A^2\xi^2 - 2A\Delta\xi 
            + m^2 - \mu^2 - \Delta^2,\\
F_2(\xi) &= B^2\xi^4 - 2Bm\xi^2 - A^2\xi^2 + 2A\Delta\xi 
            + m^2 - \mu^2 - \Delta^2.
\end{align}

For $\mu=0$, the quartic factors simplify to:

\begin{align}
F_1(\xi) &= (B\xi^2-A\xi - m - \Delta)(B\xi^2 + A\xi - m + \Delta),\\
F_2(\xi) &= (B\xi^2-A\xi - m + \Delta)(B\xi^2 + A\xi - m - \Delta),
\end{align}

This fully factorized form permits exact analytical root extraction for $\mu=0$. For generic finite chemical potential $\mu\neq0$, simple exact factorization is no longer feasible; we therefore adopt controlled analytical approximations valid in the small-$B$ limit consistent with our physical parameter regime.

\subsection{Analytical Approximations for Small $B$}

\textit{Four small roots.}-- Neglecting the subdominant $B\xi^2$ terms first isolates the low-magnitude roots governing near-boundary state decay. The four resulting approximate solutions read
\begin{align}
\xi_{1,2} &\approx \frac{\Delta \pm \sqrt{m^2 - \mu^2}}{A}, \qquad
\xi_{3,4} \approx -\frac{\Delta \pm \sqrt{m^2 - \mu^2}}{A}.\label{eq:xi34}
\end{align}
Finite $B$ only introduces tiny quantitative corrections to these small-$\xi$ values and does not alter the qualitative boundary-state decay behavior.

\textit{Four large roots.}--A second set of four large-$\xi$ roots originates from balancing the dominant $B\xi^4$ and $A^2\xi^2$ terms in the quartic polynomials, giving
\begin{equation}
\xi_{5,6} \approx \pm\frac{A}{B},\qquad \xi_{7,8} \approx \pm\frac{A}{B}.
\label{eq:xi_large}
\end{equation}
Since $B$ takes a very small numerical value in our parameter set, these roots possess vastly larger magnitudes than the small-$\xi$ modes.

\subsection{Validity conditions of the asymptotic approximation}
The preceding small-$B$ asymptotic decomposition relies on a well-ordered perturbative hierarchy, which imposes strict magnitude constraints on the neglected $B$-dependent contributions for each family of roots. We split the validity analysis for finite small-$\xi$ roots and divergent large-$\xi$ roots separately.

\textit{Validity conditions for small-$\xi$ roots}--For the low-magnitude roots given in Eq.~\eqref{eq:xi34}, the higher-order $B$-containing monomials must be subdominant relative to the leading $A^2\xi^2$ term, which requires
\begin{equation}
\left| B^2\xi^4 \right| \ll \left| A^2\xi^2 \right|, \qquad
\left| 2Bm\xi^2 \right| \ll \left| A^2\xi^2 \right|.
\end{equation}
Cancelling the common positive factor $\xi^2$ reduces the pair of inequalities to
\begin{equation}
B^2\xi^2 \ll A^2,\qquad 2|Bm| \ll A^2.
\end{equation}
Substituting the analytical form $\xi_{1,2} $ into the first inequality yields a unified single small-parameter threshold for $B$:
\begin{equation}
B \ll \min\left(\frac{A^2}{|m|},\,\frac{A^2}{\Delta \pm \sqrt{m^2 - \mu^2}}   \right).
\label{eq:valid_smallxi}
\end{equation}

\textit{Validity conditions for large-$\xi$ roots}--For the large-magnitude roots scaling as $\xi = O(1/B)$, self-consistency of the leading-order power balance requires the quartic $B^2\xi^4$ and quadratic $A^2\xi^2$ terms to dominate both the linear $2A\Delta\xi$ term and the constant offset $m^2-\mu^2-\Delta^2$. Translating this magnitude ordering into inequalities:
\begin{equation}
\frac{A^2}{B^2} \gg \frac{2A\Delta}{B}, \qquad
\frac{A^4}{B^2} \gg |m^2 - \mu^2 - \Delta^2|.
\end{equation}
Simplifying both relations yields a single sufficient condition on $B$:
\begin{equation}
B \ll \frac{A^2}{\max\bigl(|\Delta|,|m|,\sqrt{|m^2 - \mu^2 - \Delta^2|}\bigr)}.
\label{eq:valid_largexi}
\end{equation}
Equations \eqref{eq:valid_smallxi} and \eqref{eq:valid_largexi} together define the complete parameter window where our small-$B$ analytical approximations remain quantitatively reliable.

\subsection{Validation against Numerical Diagonalization}
We benchmark our approximate analytical expressions against full numerical root-finding of $\det\mathcal{H}(0,\mathrm{i}\xi)=0$. All model parameters are fixed to $m= -1 \, \mathrm{meV}$, $\mu= 0.5 \, \mathrm{meV}$, $\Delta = 1 \, \mathrm{meV}$, $A = 0.3 \, \mathrm{meV}\cdot\mu\mathrm{m}$, with a tiny quadratic coefficient $B = 1.5 \times 10^{-4} \, \mathrm{meV}\cdot\mu\mathrm{m}^2$ consistent with the small-$B$ expansion limit.
\begin{figure}[h]
    \centering
    \includegraphics[width=0.5\linewidth]{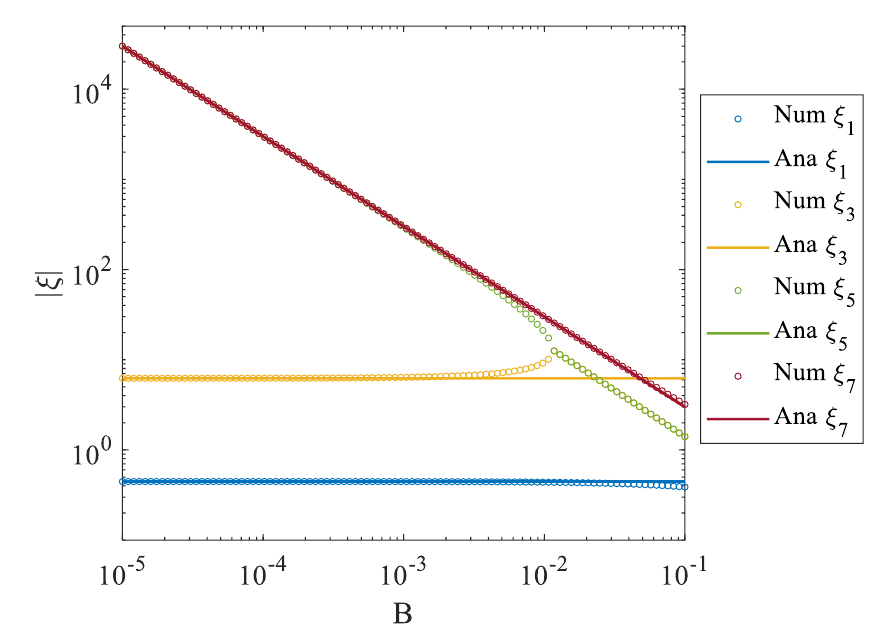}
    \caption{Comparison between numerically solved roots and the small-$B$ analytical approximations for varying $B$. Model parameters are fixed as $m= -1 \, \mathrm{meV}$, $\mu= 0.5 \, \mathrm{meV}$, $\Delta = 1 \, \mathrm{meV}$, $A = 0.3 \, \mathrm{meV}\cdot\mu\mathrm{m}$. Excellent matching is observed in the limit $B\to0$, while the analytical approximation gradually loses validity once $B \sim 10^{-2} \,\mathrm{meV}\cdot\mu\mathrm{m}^2$, violating the small-$B$ bounds derived in Sec.~II.}
    \label{fig:root_benchmark}
\end{figure}

\begin{table}[h]
\centering
\caption{Side-by-side comparison of numerically computed roots and small-$B$ analytical approximations for parameters $m= -1 \, \mathrm{meV}$, $\mu= 0.5 \, \mathrm{meV}$, $\Delta = 1 \, \mathrm{meV}$, $A = 0.3 \, \mathrm{meV}\cdot\mu\mathrm{m}$, $B = 1.5 \times 10^{-4} \, \mathrm{meV}\cdot\mu\mathrm{m}^2$.}
\begin{tabular}{c c c c c c c c c}
\hline
 & $\xi_1$ & $\xi_2$ & $\xi_3$ & $\xi_4$ & $\xi_5$ & $\xi_6$ & $\xi_7$ & $\xi_8$ \\
\hline
Numerical & 0.4465 & 0.4465 & 6.2423 & 6.2423 & $2.0136\times10^3$ & $2.0136\times10^3$ & $2.0202\times10^3$ & $2.0202\times10^3$ \\
Analytical & 0.4466 & 0.4466 & 6.2201 & 6.2201 & $2.0202\times10^3$ & $2.0202\times10^3$ & $2.0202\times10^3$ & $2.0202\times10^3$ \\
\hline
\end{tabular}
\label{tab:root_compare}
\end{table}

Table \ref{tab:root_compare} demonstrates near-perfect quantitative agreement between analytical and numerical roots when $B$ is suppressed to very small values satisfying Eqs.~\eqref{eq:valid_smallxi} and \eqref{eq:valid_largexi}. We further illustrate the systematic breakdown of the small-$B$ approximation as $B$ increases in Fig.~\ref{fig:root_benchmark}:

For $B \gtrsim 10^{-2} \,\mathrm{meV}\cdot\mu\mathrm{m}^2$, the inequalities in Eqs.~\eqref{eq:valid_smallxi} and \eqref{eq:valid_largexi} are no longer satisfied; higher-order $B$-dependent corrections can no longer be neglected, and the simple analytical formulas derived above deviate noticeably from full numerical results.

\section{Detailed Derivation of Dispersion Relation}

In the main text, we presented all edge states without explicitly specifying the normalization coefficient, which is necessary for calculating the dispersion relation. Therefore, we first calculate the normalization coefficient.

The wavefunction is given as:
\begin{equation}
\bm{\Psi}(y) = \frac{1}{\sqrt{C}} \left(c_1 \bm{v}_1 \mathrm{e}^{-\xi_1 y} + c_2 \bm{v}_2 \mathrm{e}^{-\xi_2 y} + c_7 \bm{v}_7 \mathrm{e}^{-\xi_7 y}\right).
\end{equation}

where
\begin{equation}
c_1 = \frac{1-\sigma}{2\sigma}, \qquad c_2 = \frac{1+\sigma}{2\sigma}, \qquad c_7 = -1.
\end{equation}
and
\begin{align}
\bm{v}_1 = (\sigma,-1,-\sigma,1)^{\mathrm{T}},\;
\bm{v}_2 = (\sigma,1,-\sigma,-1)^{\mathrm{T}},\label{eq:Psi12}
\bm{v}_7 = (1,1,-1,-1)^{\mathrm{T}}.\;
\end{align}
Using the normalization condition:
\begin{equation}
\int_0^\infty \mathrm{d}y \, \bm{\Psi}^\dagger(y) \bm{\Psi}(y) = 1,
\end{equation}
the normalization coefficient \( C \) is computed as:
\begin{equation}
\begin{aligned}
C=&\frac{1}{\xi_1 + \xi_2^*} \left[1 - \frac{1}{2\sigma} - \frac{1}{2|\sigma|^2} - \frac{|\sigma|^2}{2} - \frac{\sigma}{2} + \frac{1}{2\sigma^*} + \frac{\sigma^*}{2} \right]\\
+&\frac{1}{\xi_2 + \xi_1^*} \left[ 1 + \frac{1}{2\sigma} - \frac{1}{2|\sigma|^2} - \frac{|\sigma|^2}{2} + \frac{\sigma}{2} - \frac{1}{2\sigma^*} - \frac{\sigma^*}{2} \right]\\
+&\frac{\xi_1}{\xi_1^2 + |\xi_1|^2} \left[ 1 - \frac{\sigma}{2} - \frac{1}{2\sigma} + \frac{|\sigma|^2}{2} - \frac{|\sigma|^2}{2\sigma} + \frac{1}{2|\sigma|^2} - \frac{1}{2\sigma^*} \right]\\
+&\frac{\xi_2}{\xi_2^2 + |\xi_2|^2} \left[ 1 + \frac{\sigma}{2} - \frac{1}{2\sigma} + \frac{|\sigma|^2}{2} - \frac{|\sigma|^2}{2\sigma} + \frac{1}{2|\sigma|^2} + \frac{1}{2\sigma^*} \right],
\end{aligned}
\end{equation}
which simplifies to:
\begin{align}
C = \frac{2\,A\,{\left(\Delta^2 + m\,\Delta + \mu^2 \right)}}{\Delta \,{\left(\Delta^2 - m^2 + \mu^2 \right)}}.
\end{align}
Here, we have neglected terms involving \(\xi_7\), assuming that \( \xi_7 \) is very large and its contributions in denominators are negligible. Next, we calculate the perturbed energy for small \(k_x\):
\begin{equation}
\begin{aligned}
E(k_x) &= \int_0^\infty \mathrm{d}y \, \bm{\Psi}^\dagger(y) \mathcal{H}_1(k_x) \bm{\Psi}(y) \\ 
&= \frac{A k_x}{C} \frac{1}{\xi_1 + \xi_2^*} \left[ 1 - \frac{\sigma}{2} + \frac{1}{2\sigma} + \frac{\sigma^*}{2} - \frac{\sigma^*}{2\sigma} - \frac{\sigma^2}{2 |\sigma|^2} + \frac{\sigma}{2 |\sigma|^2} \right] \\
&+ \frac{A k_x}{C} \frac{1}{\xi_2 + \xi_1^*} \left[ 1 + \frac{\sigma}{2} - \frac{1}{2\sigma} - \frac{\sigma^*}{2} - \frac{\sigma^*}{2 \sigma} - \frac{\sigma^2}{2 |\sigma|^2} - \frac{\sigma}{2 |\sigma|^2} \right] \\
&+ \frac{A k_x}{C} \frac{\xi_1}{\xi_1^2 + |\xi_1|^2} \left[ 1 - \frac{1}{2\sigma} - \frac{\sigma}{2} - \frac{|\sigma|^2}{2 \sigma} + \frac{|\sigma|^2}{2 \sigma^2} + \frac{\sigma^2}{2 |\sigma|^2} - \frac{\sigma}{2 |\sigma|^2} \right] \\
&+ \frac{A k_x}{C} \frac{\xi_2}{\xi_2^2 + |\xi_2|^2} \left[ 1 + \frac{1}{2\sigma} + \frac{\sigma}{2} + \frac{|\sigma|^2}{2 \sigma} + \frac{|\sigma|^2}{2 \sigma^2} + \frac{\sigma^2}{2 |\sigma|^2} + \frac{\sigma}{2 |\sigma|^2} \right].
\end{aligned}
\end{equation}
After simplifications, this yields:

\begin{align}
E(k_x) = \frac{ \Delta \left(\Delta + m\right)}{\Delta^2 + m \Delta + \mu^2}A k_x.
\end{align}

 \twocolumngrid

\bibliographystyle{apsrev4-2}
\bibliography{Refs}
\end{document}